\documentclass[aps,pre,twocolumn,amssymb,floatfix]{revtex4}
\usepackage{graphicx}
\usepackage[hyperindex]{hyperref}

\begin{document}

\title{The criticality of the Hantavirus infected phase at Zuni}

\author{G. Abramson}
\email{abramson@cab.cnea.gov.ar}

\affiliation{Centro At\'{o}mico Bariloche, CONICET and Instituto
Balseiro, 8400 San Carlos de Bariloche, R\'{\i}o Negro, Argentina}

\affiliation{Consortium of the Americas for Interdisciplinary
Science, University of New Mexico, Albuquerque, New Mexico 87131,
USA}

\begin{abstract}
A preliminary analysis of the temporal evolution of a population
of \emph{Peromyscus maniculatus} infected with Hantavirus Sin
Nombre is made. Ecological and epidemiological parameters are
derived from the data, and they are used as inputs for the
analytical model presented in [Abramson and Kenkre, Phys. Rev. E
\textbf{66}, 011912 (2002)]. A prediction of the critical carrying
capacity and its associated critical mouse density is made, and
the time series is analyzed under the light of these. It is found
that the sporadic disappearances and reappearances of the infected
phase correspond to the bifurcation predicted by the model.
\end{abstract}


\date{\today}

\maketitle

\begin{figure}[h]
\centering
\resizebox{\columnwidth}{!}{\includegraphics{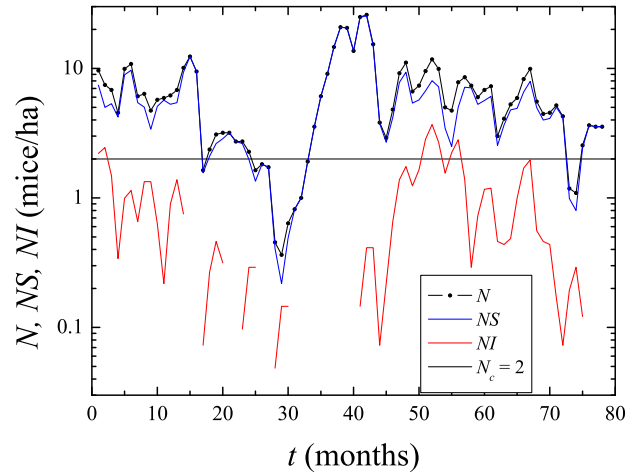}}
\caption{Population of
\emph{P. maniculatus} at the Zuni capture site (New Mexico),
from~\cite{yates02}. Total, susceptible and infected populations are shown, as
indicated in the legend. Reference~\cite{yates02} provides data for the total
and the seropositive populations, here represented as $N$ and $N_I$
respectively. The susceptible population $N_S$ is defined as $N-N_I$.}
\label{figzuni}
\end{figure}

\begin{figure}[t]
\centering
\resizebox{\columnwidth}{!}{\includegraphics{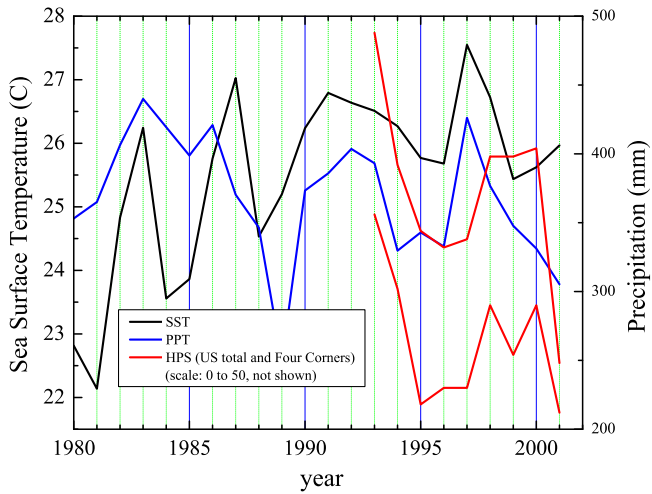}}
\caption{The connection between the environment, in particular
weather, and incidence of HPS becomes apparent in this figure,
that shows an average sea surface temperature (SST) of the East
Pacific Ocean, average precipitation (PPT) in Southwest North
America, and cases of HPS in the USA (total  and in the Four
Corners region, near Zuni). Sources: The SST has been calculated
as an average of all the East Pacific buoys of the TAO (Tropical
Atmosphere Ocean) Project (www.pmel.noaa.gov/tao); the PPT are the
"Southwest" values provided by the National Climatic Data Center
(www.ncdc.noaa.gov); HPS cases are from Centers for Disease
Control and Prevention (www.cdc.gov).} \label{weather}
\end{figure}

This report contains a first attempt to test, in a quantitative way, the
predictions of the analytical model for the Hantavirus epizootic proposed by
Abramson and Kenkre~\cite{abramson2002,abramson2003}. The analysis is based on
ecological and epidemic data published by Yates et al.~\cite{yates02},
consisting of a time series of mice populations at two sites near Zuni, New
Mexico, covering monthly a period from 1995 to 2001 (78 months). The population
data are displayed in Fig.~\ref{figzuni}, where the total, susceptible and
infected populations of \emph{Peromyscus maniculatus}, the main host of
Hantavirus Sin Nombre, are shown as $N$, $N_S$ and $N_I$ respectively. The
period does not include the outbreak of 1993, when the virus was discovered to
be the cause of the severe Hantavirus Pulmonary Syndrome (HPS). The great peak
that can be seen around month 40 in Fig.~\ref{figzuni} corresponds to a
population explosion following the Ni\~{n}o event of 1997-1998, that was in turn
followed by a small outbreak of $N_I$ and HPS in 1999 (see Fig.~\ref{weather}).
Observe also that the vertical scale is logarithmic to facilitate the
visualization of small values. It can be seen that $N_I$ drops to 0 several
times, and the corresponding line is discontinuous at these values because of
the logarithm (not because of a lack of data). These are the sporadic
disappearance of the infected phase, reported
in~\cite{mills99,parmenter99,yates02}, and that are accounted for as
bifurcations controlled by the environmental carrying capacity in the model of
Ref.~\cite{abramson2002}. The assumptions of the present analysis are big but
the results seem encouraging and support the analytical results of
Ref.\cite{abramson2002}.

The basic model, mean-field-like and not extended in space, for
the dynamics of the mice populations is:
\begin{eqnarray}
\frac{dN_S}{dt} &=& b N-cN_S-\frac{N_S N}{K} - aN_S N_I,\\
\frac{dN_S}{dt} &=& -cN_I -\frac{N_I N}{K}+aN_S N_I,
\end{eqnarray}
where $b$ is the birth rate, $c$ is the death rate, $a$ is the
contagion rate, and $K$ is the carrying capacity. The interested
reader may consult Refs.~\cite{abramson2002,abramson2003} for a
detailed discussion on the motivation and the implications of this
model, as well as for a discussion of infection waves observed in
the spatially extended model.

Let us make certain assumptions that will allow us to derive
values for the parameters from the data. The first assumption is
that $b$, $c$ and $a$ are constant, determined by biological
properties of the agents involved, and independent of anything
that is changing in the system and that makes the populations grow
and decline. This role is reserved solely for $K=K(t)$. This is,
indeed, the same assumption made in
Refs.~\cite{abramson2002,abramson2003}, to analyze the bifurcation
and to simulate a time dependent scenario.

A further assumption is to use simplified models to extract the
parameters from restricted regions of the graphs. For example,
observe the best visible population explosion of $N$, from $t=30$
to 40. It is a very well defined exponential, and it strongly
suggests the use of
\begin{equation}
\frac{dN}{dt}= (b-c) N, \label{eqbc}
\end{equation}
that is, the linear part of the logistic model for $N$. So, the
simplest way to calculate $b-c$ is to suppose an exponential
growth in all the total population increases.

If the processes of natural death and contagion can be supposed
decoupled to some extent, we can extract $a$ from the explosions
of $N_I$ (contagion is the only source of $N_I$) and $c$ from its
decays (since $-c N_I$ is the linear part of the sink of $N_I$).

So, the third assumptions is that the decays of $N_I$ can be
supposed exponential:
\begin{equation}
\frac{dN_I}{dt} = -c N_I, \label{eqc}
\end{equation}
whence the parameter $c$ can be obtained.

Now, the only source of $N_I$ is the infection, so in the
explosions of $N_I$ we suppose the simplified equation:
\begin{equation}
\frac{dN_I}{dt} = a N_I N_S, \label{eqa}
\end{equation}
that constitutes the fourth assumption.

The three Eqs. (\ref{eqbc}), (\ref{eqc}) and (\ref{eqa}) can be
solved (this last only approximately) and the solutions fitted to
the data. The exponential growth of $N$ is the nicest, because
there are several relatively broad regimes of such behavior. The
two other processes, decay and infection, are less clear cut,
because both are derived from the same $N_I(t)$, and also because
the regimes are narrower.

The results are the following:
\begin{equation}
b-c = 0.21,~~~~c = 0.75,~~~~ a = 0.77, \label{parameters}
\end{equation}
in units of month$^{-1}$ ($c$ and $b-c$) and
(mice$\times$month)$^{-1}$ ($a$). It is imperative to take this
values \emph{cum grano salis}. Besides the assumptions mentioned
above, there are further uncertainties. In the average $a$, for
example, a value of 17 was discarded because it is so bigger than
all the others that is inevitable to suspect some artifact of the
time series that, at this point, cannot be clarified. Let us carry
on, nevertheless, and from (\ref{parameters}) and
Ref.~\cite{abramson2002} conclude that
\begin{equation}
K_c=\frac{b}{a(b-c)}\approx 6\,\, \mbox{mice}\times\mbox{month}.
\end{equation}

This value of $K_c$ is the main quantitative result of the present
analysis. It is a prediction of the model, based on the numerical
values of the parameters as can be estimated from the dynamics of
the populations. How does it compare with features of the time
series? If we observe that the equilibrium solution of the
logistic equation satisfied by $N$, when $K$ is constant, is
$N^*=K(b-c)$, there is a ``critical density'' $N_c=K_c (b-c)$ with
which the actual density can be compared. When not growing or
decaying, $N(t)$ will be more or less at equilibrium during the
periods that $K(t)$ remains constant. $N_c=2$ mice, approximately
corresponding to the calculated parameters, is shown as a black
horizontal line in Fig.~\ref{figzuni}.

Let us analyze the time series under the light of this result. At
$t<15$ the population is above critical, and there is,
correspondingly, a positive infected phase. During this regime, it
is conceivable that $K(t)$ has some time dependence, but that it
keeps it above critical. Then, around $t=15$, $K(t)$ drops to some
value below critical, and so does $N^*$, and in consequence $N(t)$
drops, trying to reach an equilibrium which is now below $N_c$.
The drop is not monotonous, there seems to be some discrete steps.
What is the result? Shortly after the decline of $N$ begins (the
indication that $K$ has gone subcritical) the infected phase
begins to disappear sporadically. A few infected mice may be
entering by migration, or marginal susceptible mice might be being
infected, but it is clear that the infection is disappearing from
the site. After $t=30$ a steady population explosion starts,
indicating that $K$ has increased, and the observation that the
population grows beyond $N_c$ indicates that the system is
supercritical again. A recovering of the infected phase is to be
expected. It takes time, however, just as in the model (see Fig. 2
in Ref.\cite{abramson2002}), and not before $t=40$ do we see a
positive $N_I$ again. After this, the population remains above
critical, so $K(t)$ must be critical most of the time, and the
infection persists. A brief excursion of $N$ below $N_c$ at
$t\approx 70$ might be the beginning of a new extinction event,
and indeed $N_I$ reaches its lowest values since $t=40$. But
shortly after this the time series ends and the analysis can not
be carried further. Observe that the drop in $N$ (so in $K$) takes
place in 2001, a year that was particularly dry, as can be seen in
the precipitation data in Fig.~\ref{weather}.

In summary, even if there are a number of uncertainties, the
analysis fits nicely in the picture given by the model of
Ref.~\cite{abramson2002}, namely that there is a critical point
and that the system lives close to it. It encourages to attempt a
more detailed study, in particular involving bigger data sets of
the same system.

\bigskip\noindent
\textbf{\centerline{ACKNOWLEDGMENTS}}

The author acknowledges support from Fundaci\'{o}n Antorchas (Argentina) and from
CONICET (PEI 6482). Discussions with V. M. Kenkre, J. Salazar, F. Koster, B.
Parmenter and T. Yates have been invaluable in this study.

\end{document}